\documentclass[manuscript]{emulateapj}

\shorttitle{Sunspot Light Walls Suppressed by Nearby Brightenings} \shortauthors{Yang et al.}

\journalinfo{Accepted for publication in ApJL}
\submitted{Accepted for publication in ApJL}

\begin{document}

\title{Sunspot Light Walls Suppressed by Nearby Brightenings}

\author{Shuhong Yang\altaffilmark{1,2}, Jun Zhang\altaffilmark{1,2},
Robertus Erd\'{e}lyi\altaffilmark{3,5}, Yijun Hou\altaffilmark{1,2},
Xiaohong Li\altaffilmark{1,2}, and Limei Yan\altaffilmark{4}}

\altaffiltext{1}{CAS Key Laboratory of Solar Activity, National
Astronomical Observatories, Chinese Academy of Sciences, Beijing
100012, China; shuhongyang@nao.cas.cn}

\altaffiltext{2}{College of Astronomy and Space Sciences, University
of Chinese Academy of Sciences, Beijing 100049, China}

\altaffiltext{3}{Solar Physics and Space Plasma Research Centre,
School of Mathematics and Statistics, University of Sheffield, Hicks
Building, Hounsfield Road, Sheffield S3 7RH, UK}

\altaffiltext{4}{Key Laboratory of Earth and Planetary Physics,
Institute of Geology and Geophysics, Chinese Academy of Sciences,
Beijing 100029, China}

\altaffiltext{5}{Department of Astronomy, E\"otv\"os Lor\'and University,
Budapest, P.O.Box 32, H-1518, Hungary}

\begin{abstract}

Light walls, as ensembles of oscillating bright structures rooted in
sunspot light bridges, have not been well studied, although they are
important for understanding sunspot properties. Using the
\emph{Interface Region Imaging Spectrograph} and \emph{Solar
Dynamics Observatory} observations, here we study the evolution of two
oscillating light walls each within its own active region (AR). The
emission of each light wall decays greatly after the appearance of
adjacent brightenings. For the first light wall, rooted within AR
12565, the average height, amplitude, and oscillation period
significantly decrease from 3.5 Mm, 1.7 Mm, and 8.5 min to 1.6 Mm,
0.4 Mm, and 3.0 min, respectively. For the second light wall, rooted
within AR 12597, the mean height, amplitude, and oscillation period
of the light wall decrease from 2.1 Mm, 0.5 Mm, and 3.0 min to 1.5
Mm, 0.2 Mm, and 2.1 min, respectively. Particularly, a part of the
second light wall becomes even invisible after the influence of nearby
brightening. These results reveal that the light walls are
suppressed by nearby brightenings. Considering the complex magnetic
topology in light bridges, we conjecture that the fading of light
walls may be caused by a drop in the magnetic pressure, where flux
is cancelled by magnetic reconnection at the site of the nearby brightening.
Another hypothesis is that the wall fading is due
to the suppression of driver source (\emph{p}-mode oscillation), resulting from
the nearby avalanche of downward particles along reconnected brightening
loops.

\end{abstract}


\section{INTRODUCTION}

Light bridges are bright structures deeply anchored in the
convection zone, and they are often accounted for the incompletely
suppressed convection (Sobotka et al. 1993; Borrero \& Ichimoto
2011; Lagg et al. 2014). The magnetic field in a light bridge is
mostly much weaker than the neighboring umbra (Rueedi et al. 1995;
Jur{\v c}{\'a}k et al. 2006; Sobotka et al. 2013).
Analysing images obtained in the 1600 {\AA}
ultraviolet (UV) channel of the \emph{Transition Region and
Coronal Explorer}, Berger \& Berdyugina (2003) found persistent
brightness enhancements over a light bridge. In some other studies,
more dynamic brightenings and surges were observed in the lower
atmosphere above sunspot light bridges (Asai et al. 2001; Shimizu et
al. 2009; Tian et al. 2014; Louis et al. 2014; Toriumi et al. 2015a,
b; Robustini et al. 2016; Song et al. 2017).

Combining observations of AR 12192 made by the
\emph{New Vacuum Solar Telescope} (Liu et al. 2014) and
\emph{Interface Region Imaging Spectrograph} (\emph{IRIS}; De
Pontieu et al. 2014), Yang et al. (2015)
found an ensemble of oscillating bright structures rooted in a light
bridge and named it \emph{light wall}. The light wall, especially
the wall top, is much brighter than the surroundings. Yang et al.
(2015) suggested that the light wall oscillations are caused by the
leakage of \emph{p}-modes from the sub-photosphere. Bharti (2015)
also noted that a wave phenomenon seems to be responsible for the
coherent behavior of neighbouring oscillating structures above
the light bridge. Afterward, a survey of seven-month \emph{IRIS}
observations by Hou et al. (2016a) reveals that most light walls are
rooted in light bridges. Recently, Zhang et al. (2017) analyzed
\emph{IRIS} spectral data of a light wall which also exhibits
pronounced oscillations in the height of the light wall.
They deduced from the blue- and red-shifted Doppler
signals that the oscillations are likely caused by shocked
\emph{p}-mode waves originated from the sub-photosphere.

Last but not least, Yang et al. (2016) found that, when falling material reached
the base of a light wall, the height and brightness of the light
wall increased, implying that the light wall can be enhanced by
external disturbance. Different from the light wall enhancement, we
report in the present Letter an unusual and puzzling phenomenon, i.e., light
walls can be significantly weakened due to the suppression of nearby
brightenings observed by \emph{IRIS} and the \emph{Solar Dynamics
Observatory} (\emph{SDO}; Pesnell et al. 2012).

\section{OBSERVATIONS AND DATA ANALYSIS}

\begin{figure*}
\centering
\includegraphics
[bb=59 312 528 520,
clip,angle=0,width=\textwidth]{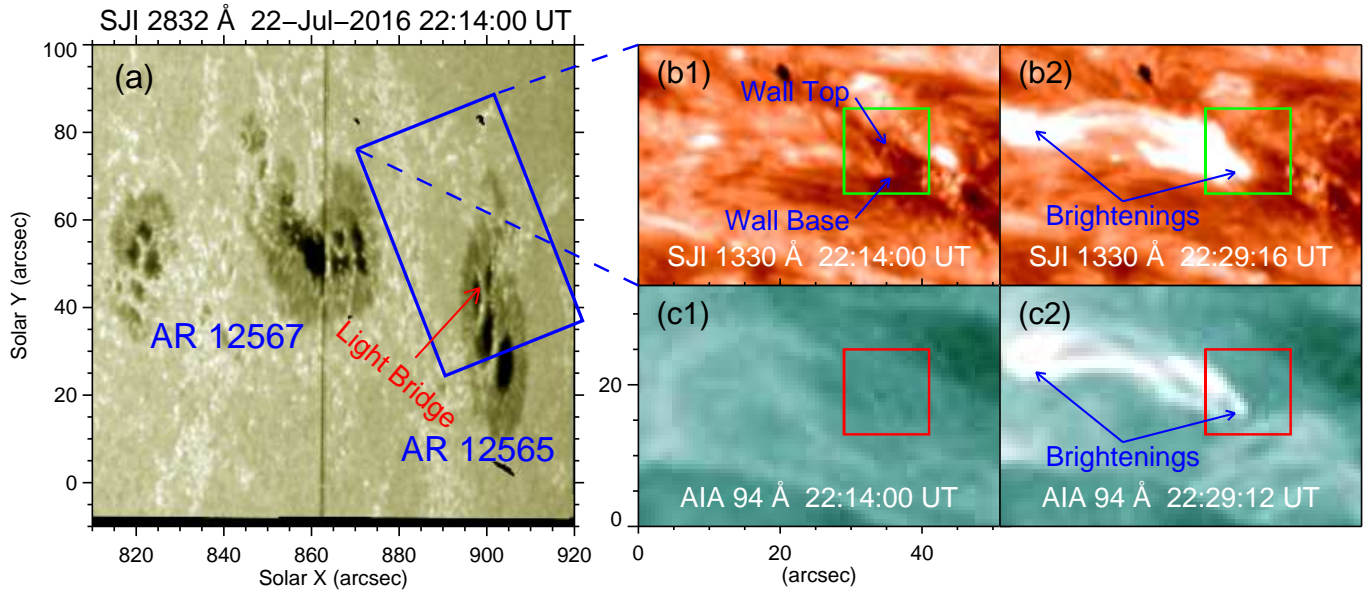}
\caption{Panel (a): \emph{IRIS}/SJI 2832 {\AA} image taken on 2016
July 22. The blue rectangle outlines the FOV of panels (b1)-(c2),
and the red arrow denotes a light bridge within a sunspot of AR
12565. Panels (b1)-(b2): SJI 1330 {\AA} images showing a light wall
rooted in the light bridge and the nearby brightenings. Panels
(c1)-(c2): \emph{SDO}/AIA 94 {\AA} images displaying the EUV
appearance corresponding to panels (b1) and (b2) (see also the
animation). The blue arrows in panel (b1) indicate the top and base
of the light wall, and the blue arrows in panels (b2) and (c2)
denote the brightening loop set. The squares outline the FOV of
Figures 2(a1)-(b3). \protect\\An associated animation (Movie1.mp4) of this
figure is available. \label{fig}}
\end{figure*}

\begin{figure*}
\centering
\includegraphics
[bb=128 198 438 646,
clip,angle=0,width=0.65\textwidth]{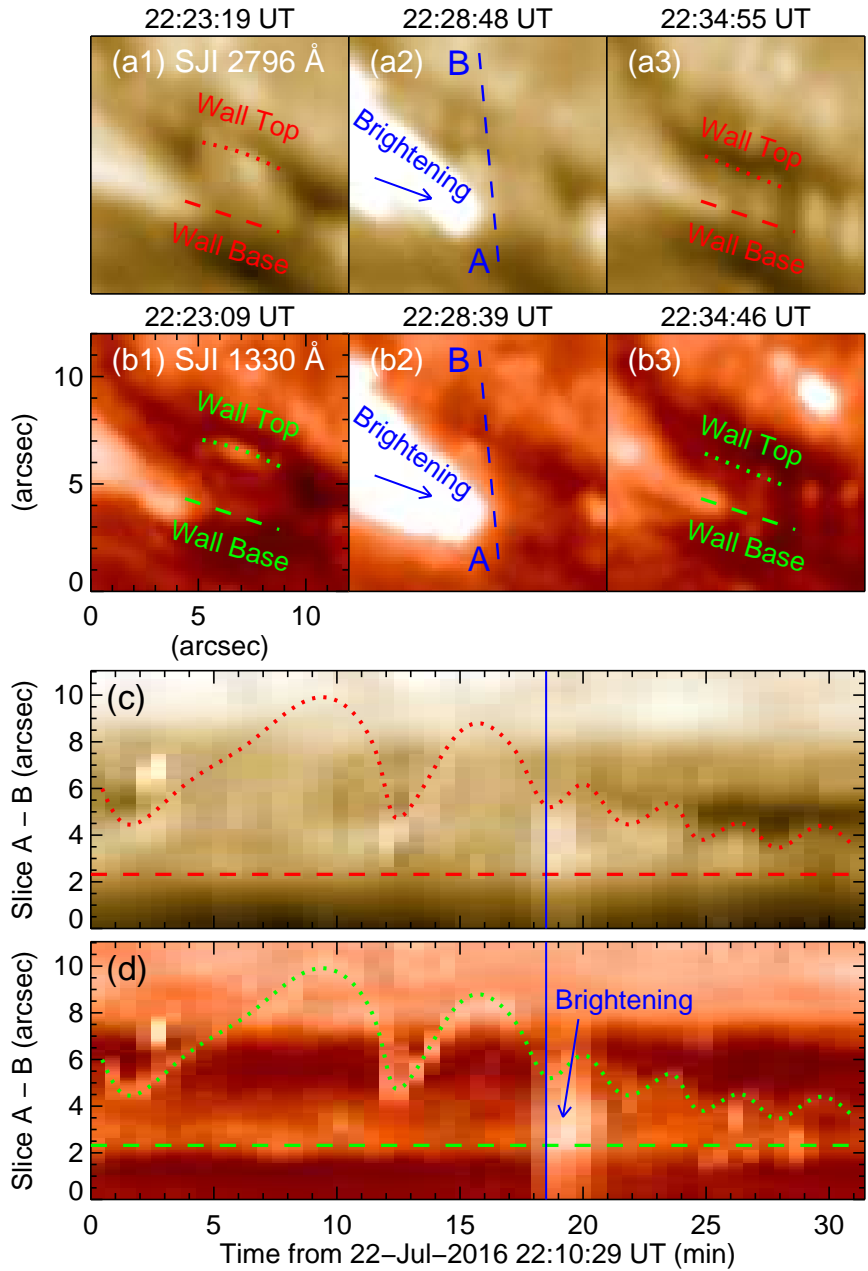}
\caption{Panels (a1)-(a3): SJI 2796 {\AA} images showing the
evolution of the light wall under the influence of the brightening
(marked by the blue arrow). Panels (b1)-(b3): Similar to panels
(a1)-(a3), but in 1330 {\AA} passband (see also the animation). The
red/green dotted and dashed curves delineate the wall top and wall
base, respectively. Panels (c)-(d): Time-distance plots derived
along slice ``A--B" marked by the blue dashed lines in panels (a2)
and (b2). The dotted curves and dashed lines outline the positions
of the wall top and wall base, respectively. The vertical lines mark
the moment of time when the brightening began to affect the light
wall. \protect\\An associated animation (Movie2.mp4) of this figure is
available. \label{fig}}
\end{figure*}

\begin{figure*}
\centering
\includegraphics
[bb=55 328 528 504,
clip,angle=0,width=1\textwidth]{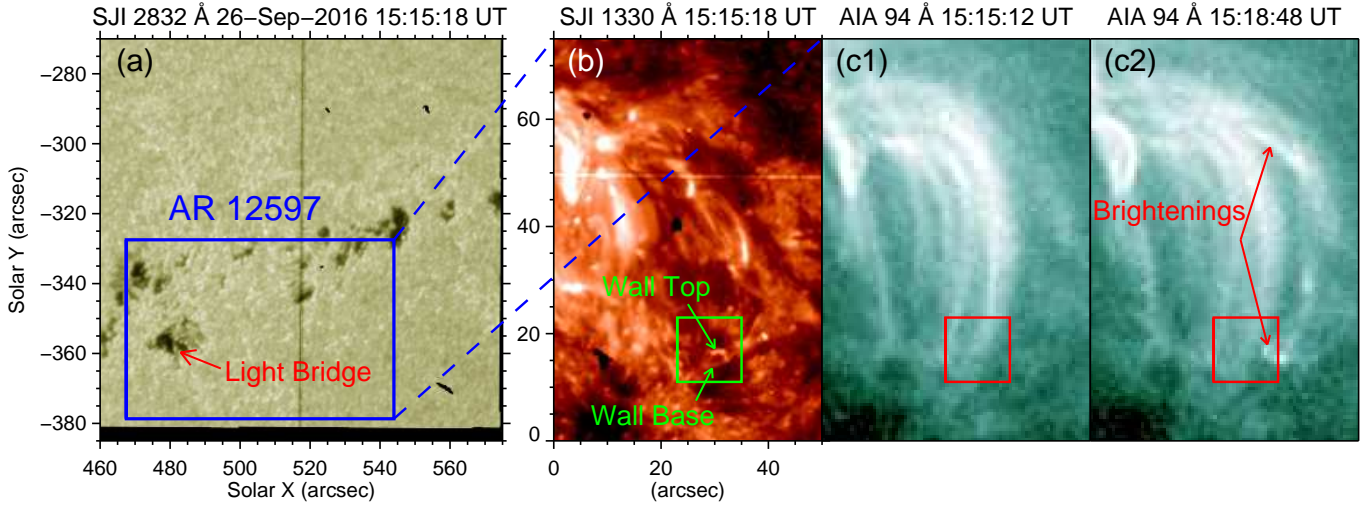}
\caption{Panel (a): SJI 2832 {\AA} image observed on 2016 September
26. The blue rectangle outlines the FOV of panels (b)-(c2), and the
red arrow denotes a light bridge within a sunspot of AR 12597. Panel
(b): SJI 1330 {\AA} image displaying the light wall rooted in the
light bridge. Panels (c1)-(c2): AIA 94 {\AA} images showing the
coronal appearance before and after the brightening of a set of
loops (marked by the red arrows; see also the animation). The
squares outline the FOV of Figure 4. \protect\\An animation
(Movie3.mp4) relevant to this figure is available. \label{fig}}
\end{figure*}

\begin{figure*}
\centering
\includegraphics
[bb=50 267 539 559,
clip,angle=0,width=0.85\textwidth]{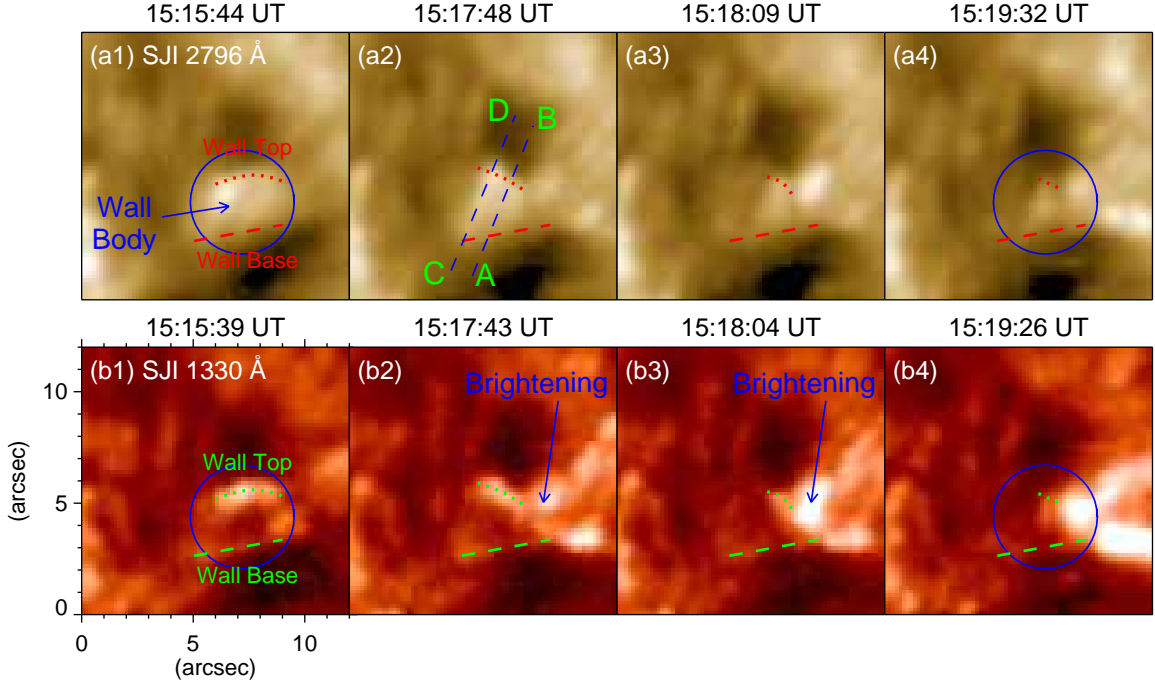}
\caption{Panels (a1)-(a4): SJI 2796 {\AA} image sequence showing the
evolution of the light wall (encompassing area outlined by the blue
circles). Panels (b1)-(b4): Similar to panels (a1)-(a4), but in 1330
{\AA} passband (see also the animation). The red/green dotted and
dashed curves mark the wall top and wall base, respectively. The
arrows in panels (b2) and (b3) denote the brightening close to the
light wall. The blue dashed lines ``A--B" and ``C--D" in panel (a2)
mark the positions where the time-distance diagrams shown in Figure
5 are obtained. \protect\\An animation (Movie4.mp4) relevant to this figure
is available. \label{fig}}
\end{figure*}

\begin{figure*}
\centering
\includegraphics
[bb=79 80 495 742,
clip,angle=0,width=0.6\textwidth]{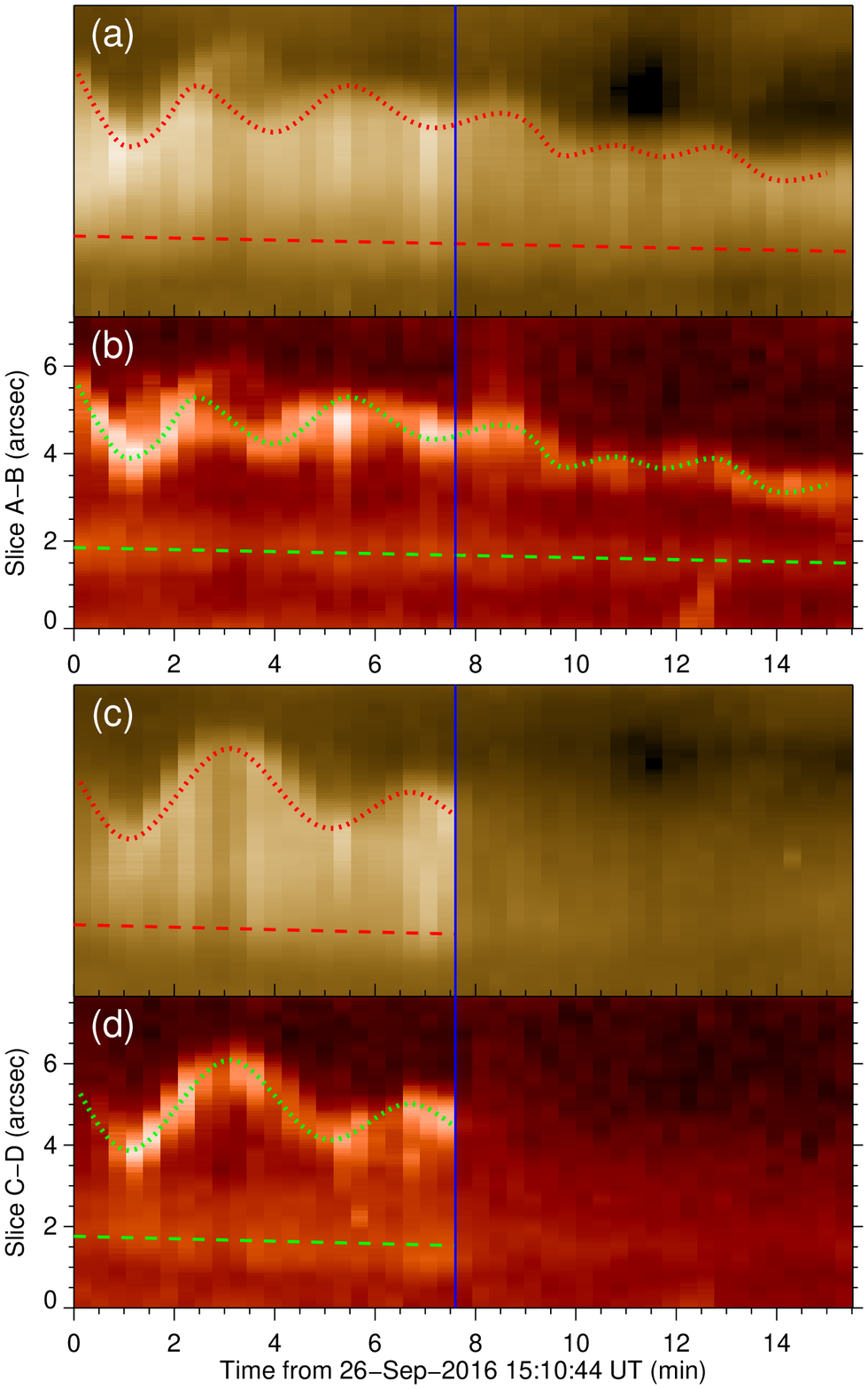}
\caption{Panels (a)-(b): Time-distance plots derived from SJI 2796
{\AA} and 1330 {\AA} images along slice ``A--B" marked in Figure 4.
Panels (c)-(d): Similar to panels (a) and (b), but along slice
``C--D." The dotted curves mark out the top of the light wall, and
the dashed lines approximate the wall base. The vertical lines
indicate the moment of time when the light wall began to be
suppressed by the nearby brightening. \label{fig}}
\end{figure*}

We study two light walls suppressed by nearby brightenings. The
first event (Event 1) was observed by \emph{IRIS} from 22:04:13 UT
on 2016 July 22 to 02:08:33 UT on July 23 with a cadence of 37 s.
The second one (Event 2) was observed also by \emph{IRIS} from 12:01:52
UT to 19:30:10 UT on 2016 September 26 with a cadence of 21 s. For
each event, there are four series of images obtained with Slit-Jaw
Imager (SJI) in 2832 {\AA}, 2796 {\AA}, 1330 {\AA}, and 1400 {\AA}
channels. These SJI images have a pixel size of 0.{\arcsec}333 and a
field-of-view (FOV) of 120{\arcsec} $\times$ 119{\arcsec}. The
Atmospheric Imaging Assembly (AIA; Lemen et al. 2012) on-board
\emph{SDO} monitors the Sun in ten (E)UV lines with a pixel size
of 0$\arcsec$.6 and cadence of (12)24 s. For these two events,
we mainly focus on two sequences of AIA 94 {\AA} images in order to study the
loop brightenings nearby the light walls. To coalign the \emph{IRIS}
images with the AIA intensity maps, we use concurrently taken continuum images from the
Helioseismic and Magnetic Imager (HMI; Scherrer et al. 2012) on-board \emph{SDO}.
Each sequence of the \emph{SDO} data are processed to Level 1.5 by
applying the standard routine aia\_prep.pro and then differentially
rotated to a reference time. Thus the AIA and HMI images are
automatically aligned. Next, we coalign the \emph{IRIS} 2832 {\AA}
data with the HMI intensity maps using the cross-correlation
method (setpts.pro available as part of the SSWIDL software tree).

\section{RESULTS}

The first light wall is rooted in a light bridge (denoted by the red
arrow in Figure 1(a)) within AR 12565, which is located at the west
of AR 12567 and close to the west limb of the solar disk. At
22:14:00 UT, the light wall, especially the wall top and wall base
(pointed by two arrows in panel (b1)) can be identified in the SJI
1330 {\AA} image. At the left side, nearby the light wall, a set of
loops with length of about 25 Mm brightened, which were quite
conspicuous at 22:29:16 UT (the footpoints of these loops are
marked by arrows in panel (b2)). Since the appearance of the
light wall and nearby brightenings in SJI 1400 {\AA} images are
similar to that in 1330 {\AA} images, we do not show the 1400 {\AA}
images here. In the simultaneous AIA image, the loops were very
bright in the EUV 94 {\AA} line (panel (c2); also see the animation
of Figure 1). The loop brightening corresponds to a B5.6
flare with the peak at 22:30 UT, according to the
\emph{Geostationary Operational Environmental Satellite} soft x-ray flare classification.
However, the loops were invisible in 94 {\AA} 15 min ahead (see panel (c1)).

Figures 2(a1)-(a3) and the associated animation show the light wall
in SJI 2796 {\AA} before, during, and after the influence of the
nearby brightening, respectively. The corresponding appearances in
SJI 1330 {\AA} are presented in panels (b1)-(b3). The top and base
of the light wall are identified from both the 1330 {\AA} and 2796
{\AA} images. At 22:23 UT, the projected height (the distance
between the wall top and wall base) was about 2.1 Mm (see panels
(a1) and (b1)). The nearby brightening extended to the light wall
from the left side and then impacted on the wall base (panels (a2)
and (b2)). After that, the light wall became much fainter and less
pronounced, both in emission and height (panels (a3) and (b3)). To
study the fading of the light wall, we derive two time-distance
plots (see panels (c) and (d)) from two sequences of SJI 2796 {\AA}
and 1330 {\AA} images along slice ``A--B" (marked by blue dashed
lines in panels (a2) and (b2)). From analysing the time-distance
plots in panels (c) and (d), we conclude that the wall top moved
upward and downward successively, indicating an oscillatory pattern
in the height of the light wall. At the time around 22:29 UT (marked
by the vertical lines), the brightening nearby the wall (denoted by the blue arrow)
began to affect the light wall. The average height of the light wall
before this influence was about 3.5 Mm. Due to the influence caused
by the brightening, the mean height of the light wall decreased to
1.6 Mm. The average amplitude of the associated oscillation also decreased, from
1.7 Mm to 0.4 Mm. In addition, the oscillation period of the light
wall has also changed. In the 18.5 min interval preceding
the interaction with the brightening, the mean period of oscillation
was about 8.5 min, and after the interaction, it has dropped down to
about 3.0 min in the following 13 min interval.

The second light wall is located within the emerging AR 12597 (see
Figure 3(a)). We note that, there is another light bridge (pointed
by the red arrow in panel (a)) in one sunspot of the AR. In the SJI
1330 {\AA} image (panel (b)), a light wall rooted in the light
bridge can be clearly identified, and its top and base are marked by
two green arrows. However, the light wall was almost invisible in
the AIA 94 {\AA} image at 15:15:12 UT (panel (c1)). Several minutes
later, a set of coronal loops with an average length of about 45 Mm
brightened, i.e., that this is an estimated length, as highlighted by
the arrows in panel (c2) (also see the animation of Figure 3).
The loop brightening corresponds to a B2.5 flare with the peak at 15:20 UT.
We note that the loop set itself connects to the opposite polarity
fields of the AR, and, one of its ends appears as brighter points
(marked by the lower arrow), located nearby the base of the light wall.

Figure 4(a1) shows that the light wall (outlined by the blue circle)
is brighter than the surrounding area observed in SJI 2796 {\AA}.
In the SJI 1330 {\AA} line, the emissions of the wall top and wall
base are much higher than those of the wall body and the surrounding region
(see panel (b1)). The right side nearby region of
the light wall brightened and was very appealing at 15:17:43 UT in
SJI 1330 {\AA}, as indicated by the arrow in panel (b2).
Then, the brightening became more violent and affected the light
wall at 15:18:04 UT, as shown in panels (a3) and (b3). The left part
of the light wall (overlaid by slice ``C-D") almost disappeared, and
only the right part (overlaid by slice ``A-B") remained (also see
the animation of Figure 4). One and a half minutes later, the right
part of the light wall became more faint in both 2796 {\AA} and 1330
{\AA} (see panels (a4) and (b4)).

In order to study the change of the light wall and to explore the
corresponding cause(s) behind, we construct the time-distance plots along
slices ``A--B" and ``C-D" (marked in Figure 4(a2)), which cross the
right and left parts of the light wall, respectively. Figures
5(a) and 5(b) are the time-distance plots obtained along slice
``A--B" from SJI 2796 {\AA} and 1330 {\AA}, respectively. We
can see that, in the 7.5 min interval before 15:18 UT
(the start time of brightening marked by the vertical line), the
light wall (exactly the right part of the wall) was very bright,
and its mean height was about 2.1 Mm. The mean amplitude and
oscillation period were about 0.5 Mm and 3.0 min, respectively.
After the influence caused by the brightening, the light
wall became less prominent, as e.g., the wall height decreased to
about 1.5 Mm; the mean amplitude became 0.2 Mm and the oscillation
period dropped down to around 2.1 min in the following 8 min.
Along slice ``C--D", two time-distance plots, derived from the
SJI 2796 {\AA} and 1330 {\AA} images, are presented in Figures
5(c) and 5(d), respectively. The light wall (in particular, the
left part of the wall) before 15:18 UT appeared as a bright
oscillating structure with the oscillation period of about 4 min.
Finally, after 15:18 UT, the light wall suddenly disappeared.

\section{CONCLUSIONS AND DISCUSSION}

Using \emph{IRIS} and \emph{SDO} multi-wavelength observations,
we studied here two oscillating light walls within two active regions. Due
to the influence of nearby brightenings, the brightness of each
light wall decayed greatly. For the first light wall, rooted within
AR 12565, the average height, amplitude, and oscillation period
significantly decreased from 3.5 Mm, 1.7 Mm, and 8.5 min to 1.6 Mm,
0.4 Mm, and 3.0 min, respectively. For the second light wall, rooted
within AR 12597, the mean height, amplitude, and oscillation period
of the right part of the light wall decreased from 2.1 Mm, 0.5 Mm,
and 3.0 min to 1.5 Mm, 0.2 Mm, and 2.1 min, respectively.
Especially, the left part of the second light wall became invisible
after the influence of nearby brightening. Our results imply that
these two light walls are suppressed by nearby brightenings.

In the study of Hou et al. (2016b), a light wall in AR 12403 was
disturbed by an eruptive flare. The light wall was suggested to
share a group of magnetic lines with the flaring loops, and the height
variation of the light wall was interpreted with the projection
effect due to the inclination changes. The upward pushing of
large-scale loops lets the light wall turn to the vertical direction, thus
resulting in the increase of the projective height of the wall.
Afterward, the formation of low-lying post-flare loops makes the
light wall seem to be lower in projection since the light wall
inclined. However, in the present study, there are only some loop
brightenings (denoted by the arrows in Figures 1(c2) and 3(c2))
instead of eruptive flares. The loop brightenings seem to be caused
by magnetic reconnection among braided field lines, which is
different from the eruptive flares with dramatic inclination changes
due to the rise of stretching lines and the formation of post-flare
loops. Therefore, the cartoons in Hou et al. (2016b) cannot be used
to explain the decreases of the height, amplitude, oscillation
period, and brightness of the light walls studied in the present
work.

Solar flares often eject material from the lower atmosphere into the
corona, and some material may fall back to the solar surface. Yang
et al. (2016) noted, when the falling material reaches the base of a
light wall, the kinetic energy is converted to thermal energy.
The heated material of the light wall let the wall itself be much brighter.
The pressure at the wall base increases, which powers the
light wall to reach greater heights.  Different from the
light wall enhancement by falling material, our results in the
present study reveal that the light walls are suppressed by nearby
brightenings. Since the height of a light wall can be determined by
the pressure at the wall base, applying the logic presented in Yang et al. (2016),
one would expect here a pressure decrease caused by the nearby
brightening. This decrease could be, e.g., due to a drop in the
magnetic pressure, where flux is cancelled by magnetic reconnection
at the site of the nearby brightening. The intermittent reconnection
may cause the changes of the light wall oscillation periods.
Another opinion may be that the decrease of light walls properties
(e.g., its height) is due to the suppression of the driver source
(\emph{p}-mode oscillation) itself, resulting from the nearby hit
of downward bulk plasma along reconnected brightening loops.
Recent studies have revealed that the magnetic fields in sunspot
light bridges are quite complex (Louis et al. 2015; Toriumi et al.
2015a, b; Yuan \& Walsh 2016). For example, Toriumi et al. (2015b)
found that, in the light bridge, the magnetic field lines are highly
inclined (almost horizontal to the solar surface in the direction along
the light bridge) and appear as serpentine or arched structures.
Thus, here we propose that, when the downward propagating bulk
plasma hits the light bridge possessing a complex magnetic topology,
it can not only affect the impact site but also influences the nearby
light wall rooted in the light bridge. However, the exact mechanism
for the light wall fading is not yet clear, and to further explore
we need more observations and (MHD) modeling.

\acknowledgments {We thank the referee for valuable comments and
Dr. Hui Tian for helpful discussion. The data are used courtesy of \emph{IRIS} and
\emph{SDO} science teams. \emph{IRIS} is a NASA small explorer mission developed
and operated by LMSAL with mission operations executed at NASA Ames Research center
and major contributions to downlink communications funded by ESA and the Norwegian
Space Centre. This work is supported by the National Natural Science Foundations
of China (11673035, 11533008, 11373004), the Youth Innovation Promotion Association
of CAS (2014043). RE is grateful to STFC (UK) and the Royal Society for the
support received in a number of grants. He also thanks the Chinese Academy of
Sciences Presidents International Fellowship Initiative, Grant No. 2016VMA045 for
support received. LM Yan is supported by National Postdoctoral Program for Innovative
Talents (grant BX201600159). }

{}

\end{document}